\documentclass[useAMS,usenatbib]{mn2e}
\usepackage{bm}
\usepackage{epsfig}
\usepackage{color}
\usepackage{amssymb}
\usepackage[fleqn]{amsmath}

\newcommand{\eq}[1]{Eq.~(\ref{#1})}

\newcommand{\fig}[1]{Fig.~\ref{#1}}

\newcommand{\lt}{\left}
\newcommand{\rt}{\right}

\newcommand{\bea}{\begin{eqnarray}}
\newcommand{\eea}{\end{eqnarray}}
\newcommand\be{\begin{equation}}
\newcommand\ee{\end{equation}}

\newcommand{\deltacs}{\delta_{\rm{CS}}}
\newcommand{\deltaSP}{\delta_{\rm{SP}}}

\newcommand{\Eeff}{E_{\rm{eff}}}

\newcommand{\bnab}{\bm{\nabla}}

\newcommand{\delperp}{\nabla_\perp}
\newcommand{\urms}{u_{\rm rms}}

\newcommand{\vu}{\bm{u}}
\newcommand{\vB}{\bm{B}}

\renewcommand{\(}{\left(}
\renewcommand{\)}{\right)}
\newcommand{\dd}{\partial}

\title[Turbulent Magnetic Reconnection in Two Dimensions]{Turbulent Magnetic Reconnection in Two Dimensions}
\author[N.\ F.\ Loureiro et al.]
{N.\ F.\ Loureiro,$^{1}$\thanks{Present address:
IPFN, Instituto Superior T\'ecnico, Lisbon 1049-001, Portugal.
E-mail: nloureiro@ipfn.ist.utl.pt}
D.\ A.\ Uzdensky,$^{2}$
A.\ A.\ Schekochihin,$^{3,4,5}$
\newauthor
S.\ C.\ Cowley$^{1,4}$ and
T.\ A.\ Yousef$^{3,4}$\\
$^{1}$EURATOM/UKAEA Fusion Association, Culham Science Centre, 
Abingdon, OX14 3DB, UK\\
$^{2}$Department of Astrophysical Sciences/CMSO, 
Princeton University, Princeton, NJ 08544, USA\\
$^{3}$Rudolf Peierls Centre for Theoretical Physics, University of Oxford, Oxford OX1 3NP, UK\\
$^{4}$Blackett Laboratory, Imperial College, London~SW7~2AZ, UK\\
$^{5}$Institut Henri Poincar\'e, Universit\'e Pierre et Marie Curie, 75231 Paris Cedex 5, France}
\begin{document}
\date{\today}
\pagerange{\pageref{firstpage}--\pageref{lastpage}}\pubyear{2009}
\label{firstpage}
\maketitle

\begin{abstract}
Two-dimensional numerical simulations of the effect of background 
turbulence on 2D resistive magnetic reconnection are presented. For sufficiently small values of the resistivity ($\eta$) and moderate values of the
turbulent power ($\epsilon$), the reconnection rate is found to have a much weaker dependence on $\eta$ than
the Sweet-Parker scaling of $\eta^{1/2}$ and is even consistent with an $\eta$-independent value.
For a given value of $\eta$, the dependence of the reconnection rate on the turbulent power exhibits a critical threshold in $\epsilon$ above which the reconnection rate is significantly enhanced.
\end{abstract}



\section{Introduction}
Magnetic reconnection is a ubiquitous plasma process whereby oppositely directed magnetic field lines break and rejoin in a different topological configuration --- see, e.g.,~\citet{Biskamp-2000} for a review.
It is probably the main mechanism behind many spectacular space- and 
astrophysical phenomena, such as sub-storms in the Earth's magnetosphere~\citep{Dungey-1961} and 
solar/stellar~\citep{Yokoyama_etal-2001} and accretion disk flares~\citep{Goodson_etal-1999}. 
Reconnection has also been suggested as a possible
mechanism behind a number of high-energy astrophysics processes,
such as magnetar giant flares~\citep{Lyutikov-2003}, $\gamma$-ray bursts~\citep{Giannios_Spruit-2006} 
and rapid TeV flares in blazars~\citep{Giannios_etal-2009}.
It is believed to lead to heating 
and non-thermal particle acceleration in astrophysical coronae~\citep{Drake_etal-2006} and 
to play an important role in magnetohydrodynamic (MHD) turbulence 
and large-scale dynamos~\citep{Brandenburg-2005}. It is also a key element in sawtooth crashes 
in tokamaks~\citep{Hastie-1998}. 

Early attempts to understand reconnection in terms of the simplest possible description of the plasma --- single-fluid resistive MHD --- led to the Sweet-Parker (SP) model~\citep{Sweet-1958,Parker-1957}, 
which predicts a reconnection rate proportional to $S^{-1/2}$, where
$S=L V_A/\eta$ is the Lundquist number, $L$ is the system size, $V_A$ is the Alfv\'en speed and $\eta$ is the plasma resistivity.
In many astrophysical environments, $S\ggg1$ [e.g., $S\sim10^{14}$ in solar flares, $S\sim10^{18}$ in the interstellar medium (ISM)], resulting in 
SP reconnection rates which are orders of magnitude slower than that
observed in the above mentioned phenomena. 
The main quest in 
reconnection research has thus been to identify the key physical mechanisms that are missing from the SP theory and that can explain these fast rates.
Following much numerical work, it is now thought
that in low-density, collisionless plasmas, nonclassical effects
[e.g., the Hall effect~\citep{Biskamp-2000} or anomalous resistivity~\citep{Malyshkin_etal-2005}] can indeed give rise to fast,
Petschek-like~\citep{Petschek-1964}, reconnection.
Fast, collisionless reconnection can only take place when the resistive width of the reconnection layer,  $\deltaSP\approx L/S^{1/2}$~\citep{Sweet-1958,Parker-1957} is small compared to the relevant kinetic scale [which is frequently the ion collisionless skin-depth~\citep{Cassak_etal-2005,Yamada_etal-2006,Uzdensky-2007}, $d_i=c/\omega_{pi}$, or else $d_e=c/\omega_{pe}$ in the case of pair plasmas].
In many astrophysical situations, e.g., the solar chromosphere, the ISM,
inside stars and accretion disks, and in the high-energy-density environments in central engines of gamma-ray bursts (GRB) and core-collapse supernovae~\citep{Uzdensky_MacFadyen-2006}, the density is so high that the 
above condition is not satisfied; for example, for the warm ionized ISM, $\delta_{SP}/d_i\sim10^{3}$ [and even larger for diffuse clouds or molecular clouds~\citep{Zweibel-1989}]; in the magnetosphere of the accretion disk
in the collapsar scenario of long GRB central engines~\citep{Uzdensky_MacFadyen-2006} $\delta_{SP}/d_e\sim 10^7$ (here $d_e$ may be more important than $d_i$
because, under some circumstances, this is expected to be
mostly pair plasma).
Therefore, the
reconnection layer is collisional and resistive MHD should apply.
The prevailing opinion
is, however, that in the resistive-MHD case the Petschek mechanism
fails~\citep{Biskamp-1986,Uzdensky_Kulsrud-2000}, and reconnection is negligibly slow, possibly described by the SP theory.
Can fast (i.e., $\eta-$independent) reconnection happen in these environments?\\
\indent Missing from the SP picture is that in Nature, most, if not all, environments where reconnection takes place are likely to be turbulent [e.g.~\citep{Retino_etal-2007}]. 
Can background turbulence significantly accelerate reconnection? The aim of this Letter is to address this question.

Turbulent reconnection has been studied previously, both theoretically~\citep{Hameiri_Bhattacharjee-1987,
Strauss-1986, Strauss-1988, LV-1999, Kim_Diamond-2001} and numerically~\citep{Matthaeus_Lamkin-1986, Fan_etal-2004, Smith_etal-2004, Watson_etal-2007, Lapenta-2008, Kowal_etal-2009}.
Given the complexity of the problem, however, a rigorous theory of turbulent reconnection is not yet available and numerical studies have been limited by resolution constraints. Thus, the role of turbulence in reconnection remains controversial.
A pioneering numerical study with decaying 2D turbulence by 
\citet{Matthaeus_Lamkin-1986} suggested that a finite level of broadband MHD fluctuations can enhance reconnection [see also~\citet{Matthaeus_Montgomery-1981}],
but the very limited computing capabilities employed
precluded a clear conclusion about the asymptotic 
regime~$\eta\rightarrow 0$.
Subsequently, in a highly influential paper, Lazarian \& Vishniac
(hereafter, LV)~\citep{LV-1999}
suggested that turbulence
can greatly accelerate reconnection by enabling multiple reconnection sites in the current sheet. The LV picture \textit{is an essentially three dimensional~(3D) process}, as the multiple reconnections mechanism they envision is topologically prohibited in two dimensions (2D). 
Recent 3D numerical simulations by \citet{Kowal_etal-2009} seem to support the LV model; however, due to computational constraints, only moderate scale separations and fairly strong turbulence (but still sub-Alfv\'enic compared to the reconnecting field) could be probed.

In 2D, the LV mechanism should not work and the expectation has been that topological constraints should prevent significant acceleration of reconnection by turbulence.
In this Letter, we present evidence contrary to this belief: our numerical results show that background turbulence can have a dramatic effect on 2D magnetic reconnection, yielding reconnection rates that exhibit a much weaker dependence on $S$ than the SP scaling of $S^{-1/2}$ and may even asymptote to an $S$-independent value as $S\rightarrow\infty$. These results call for a further theoretical effort to understand magnetic reconnection in the presence of turbulence.


\section{Numerical Setup}
Our aim is to investigate how relatively weak MHD turbulence 
affects reconnection in 2D. 
The incompressible MHD equations, with an external forcing term $\bm F$, are:
\begin{align}
&\dd_t\vu+\vu\cdot\bnab\vu =\vB\cdot\bnab\vB-
\bnab \(p+B^2/2\)+\nu\nabla^2\vu + \bm F,\\
&\dd_t\vB+\vu\cdot\bnab\vB =\vB\cdot\bnab\vu+\eta\nabla^2\vB,
\end{align}
where $\vu$ is the velocity, $\vB$ is the magnetic field, and $p$ is
the pressure. 
The pressure gradient is determined by the incompressibility condition $\bm \nabla \cdot \vu=0$ (which sets the plasma density $\rho=const$), so an energy equation is not required.
The resistivity of the plasma is denoted by 
$\eta$ and the viscosity by $\nu$.
We have normalized $p$ and $B$ by 
$\rho$ and $(4\pi\rho)^{1/2}$, respectively.
Expressing the 2D [$(x,y)$] incompressible velocity field 
($\bm\nabla\cdot\vu=0$)
in terms of the stream function $\phi$ and the solenoidal magnetic field
($\bm\nabla\cdot\vB=0$) in terms of the flux function $\psi$,
$\vu=(-\dd_y\phi,\dd_x\phi)$, $\vB=(-\dd_y\psi,\dd_x\psi)$,
we can rewrite the above equations in terms of $\phi$ and $\psi$
as follows~\citep{strauss_76}:
\bea
\label{RMHD_vort}
\dd_t\nabla^2\phi + \lt\{\phi,\nabla^2\phi\rt\} &=&
\lt\{\psi,\delperp^2\psi\rt\}+\nu\nabla^4\phi+f,\\
\label{RMHD_psi}
\dd_t\psi + \lt\{\phi,\psi\rt\} &=& 
\eta\nabla^2\(\psi-\psi_{\rm eq}\),
\eea
where the 
Poisson brackets are denoted by
$\lt\{\phi,\psi\rt\}=\dd_x\phi\dd_y\psi-\dd_y\phi\dd_x\psi$. 
These are the equations that we solve with our numerical code.
The dimensions of length are set by the box size $(L_x,L_y)=(2\pi,~2.18\pi)$ and the dimensions of magnetic field are set by the equilibrium configuration: we use a tearing-mode unstable 
equilibrium, $\psi_{\rm {eq}}=\psi_0/\cosh^2(x)$, where $\psi_0=3\sqrt{3}/4$, so that the maximum value of the equilibrium magnetic field, 
$B_0=({\rm d}\psi_{\rm eq}/{\rm d} x)_{\rm max}=1$. 
The background density is normalized to 1 and 
the dimensions of velocity are given by setting the Alfv\'en speed 
$V_A=B_0=1$. Thus, time is measured in Alfv\'enic units, $L_x/(2\pi B_0)$.
Diffusion of the equilibrium magnetic field is prevented by the addition of an external electric field to the RHS of \eq{RMHD_psi} --- the term $-\eta\nabla^2\psi_{\rm eq}$.
The externally imposed (turbulent) forcing $f=\hat z\cdot \(\bm \nabla\times \bm F\)$ (random, white
noise in time), is characterized by two parameters: $\epsilon=\left\langle \vu\cdot\bm F\right\rangle$, 
the power input per unit area and
$k_f$, the forcing wave number. Note that the forcing is applied only to the momentum equation, and does not break the frozen-flux constraint.
The equations are solved in a doubly periodic box using a pseudo-spectral, symplectic algorithm~\citep{LH_JCP08} (which can be used in either semi-implicit or explicit timestepping mode: the former is employed in the laminar simulations; the latter in the turbulent ones).
Our ultimate goal is to characterize the basic behaviour of the effective 
reconnection rate~$\Eeff$ (defined below) in the 4D parameter space: $\Eeff(\eta,\nu,\epsilon,k_f)$. However, in the present study we focus on 
only two parameters, $\eta=\nu$ (i.e., the magnetic Prandtl number $Pm \equiv \nu/\eta=1$) and~$\epsilon$.	

We start our simulations without turbulence following a standard tearing mode evolution.
After the linear~\citep{FKR}, \citet{Rutherford-1973}, and $X$-point 
collapse stages~\citep{Loureiro_etal-2005}, a thin SP current layer 
forms between two large magnetic islands. 
For our parameters, this happens when the value of the reconnected 
flux at the $X$-point is $\psi_X\approx0.4$, whereupon the 
length of the current sheet is $L\approx1$, so we define $S=LV_A/\eta=1/\eta$.
Since our intention is to focus solely on the effect of turbulence 
on resistive reconnection, we use the laminar SP configuration at $\psi_X\approx0.4$ 
as a starting point for our turbulent runs and switch on the driving 
term~$f$ in~\eq{RMHD_vort} at this stage.
Furthermore, when $\psi_X\approx0.9$, the SP stage gives way 
to saturation~\citep{Loureiro_etal-2005}. Thus, we restrict our 
analysis of the numerical data to the time interval where $0.4\lesssim\psi_X\lesssim0.9$.

Our runs are characterized by the following parameters. 
We scan the range in $\epsilon$ from $\epsilon=3\times10^{-4}$ to~$\epsilon=0.1$,
and in $\eta$ from $\eta=10^{-3}$ to~$\eta=6.5\times10^{-5}$.
The forcing wave number is $k_f=10$, giving turbulent motions with characteristic scales a few times smaller than the current sheet length ($l_f \sim 0.3$).
For $\eta=6.5\times10^{-5}$, the width of the laminar current sheet is $\deltacs\approx0.014$, so we have a reasonably good scale separation in these simulations.
Resolutions up to $8192^2$ were used. Such resolutions would not have been possible in 3D.
Convergence tests were performed on selected runs to ensure that increasing the resolution did not change the results.
	
\section{Results}
\begin{figure}	
\centering
	\hspace*{-.6cm}
	\includegraphics[width=10.cm]{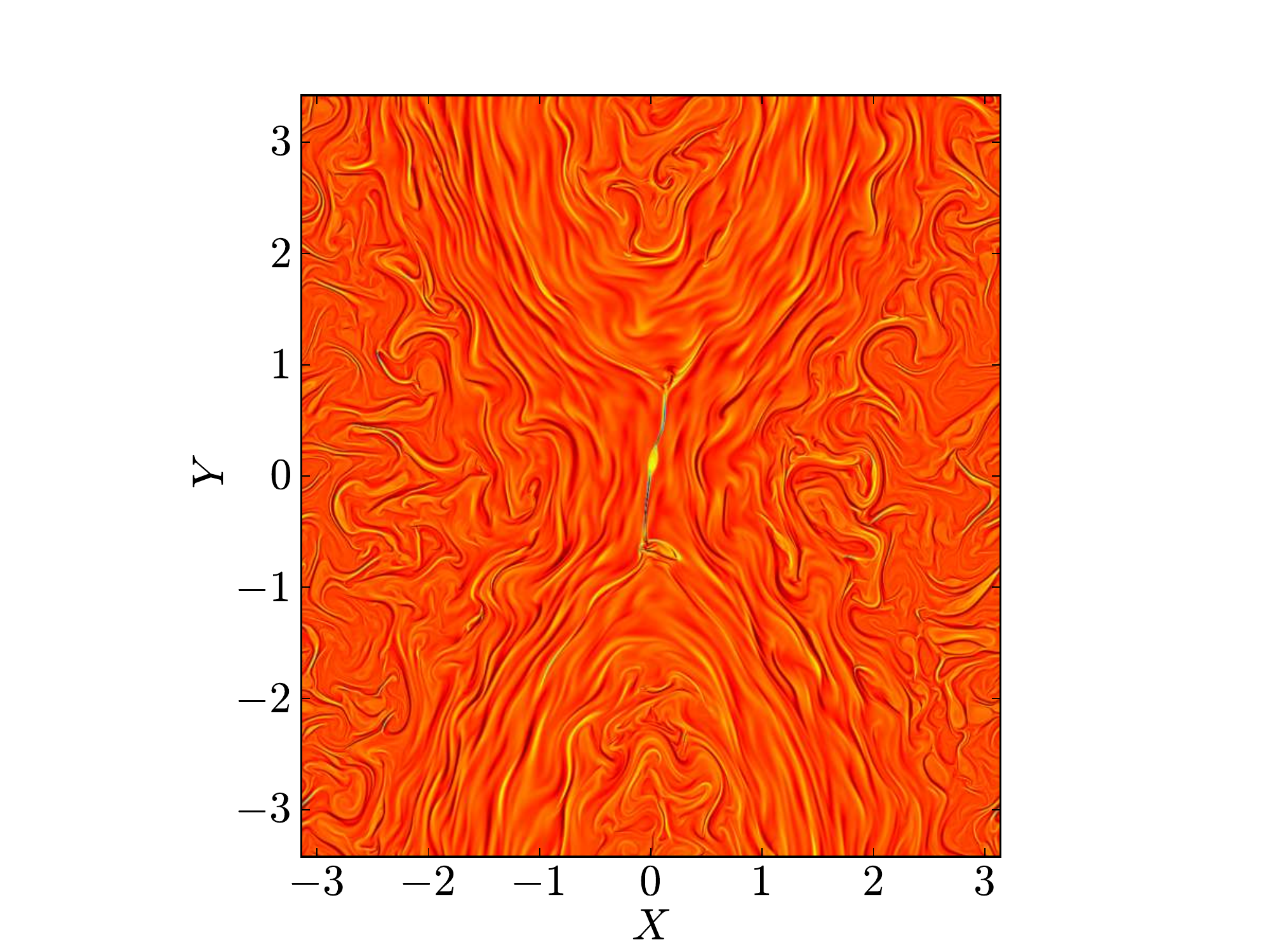}
	\caption{Magnetic reconnection in the presence of background turbulence.  Contour plot of current density $j_z=\delperp^2\psi$ at $t=735$ from the run with
parameters $\eta=6.5\times10^{-5}$, $k_f=10$ and $\epsilon=0.01$ (cf. blue dotted line of~\fig{timetrace}).
Run resolution is $8192^2$.}
	\label{global_snap}
\end{figure}

\begin{figure}
\centering
\vspace{-.5cm}
	\begin{tabular}{c}
		\hspace*{-0.8cm}
		\includegraphics[width=10cm]{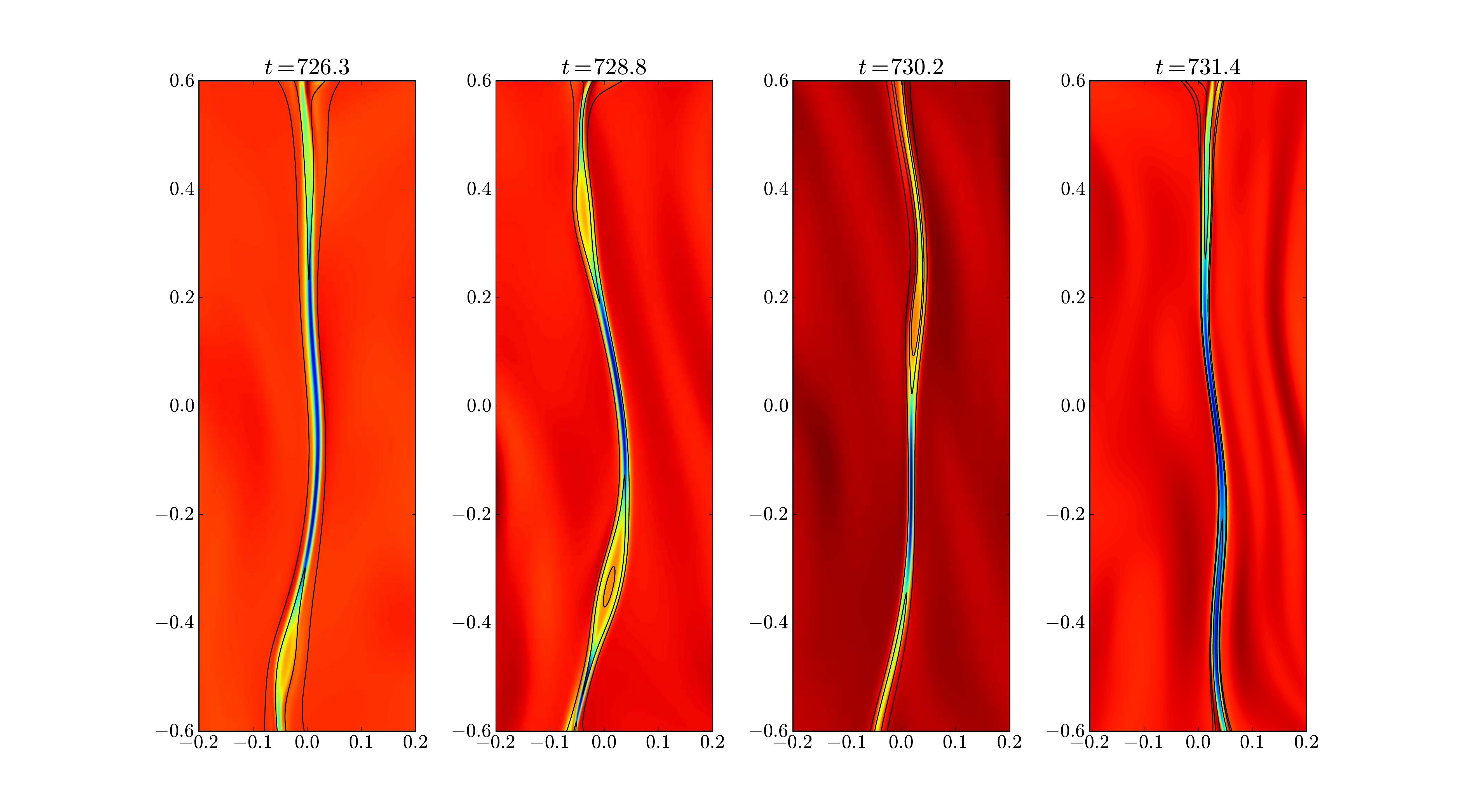}
		\vspace{-.4cm}\\
		\hspace*{-0.8cm}
		\includegraphics[width=10cm]{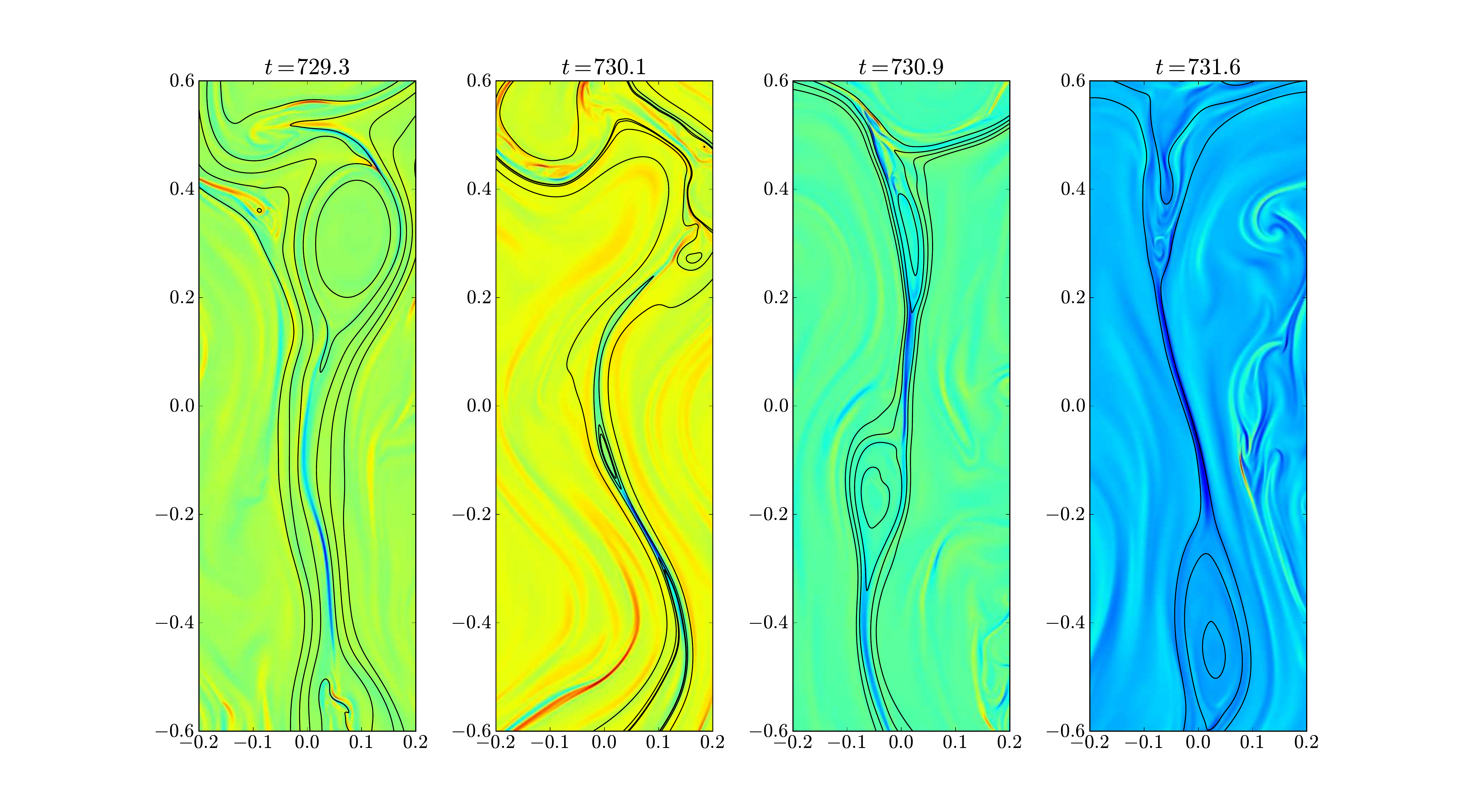}
	\end{tabular}	
\caption{Contour plots of the current density at different times for the runs with $\eta=6.5\times10^{-5}$, $k_f=10$ and $\epsilon=0.01$ (top row, cf. blue dotted line of~\fig{timetrace}) and $\epsilon=0.1$ (bottom row, cf. red full line of~\fig{timetrace}). Resolution is $8192^2$. Selected contour lines of $\psi$ are overplotted. 
Snapshots zoom in on the current sheet region.}
\label{contours}
\end{figure}

\begin{figure}
		\includegraphics[height=6cm]{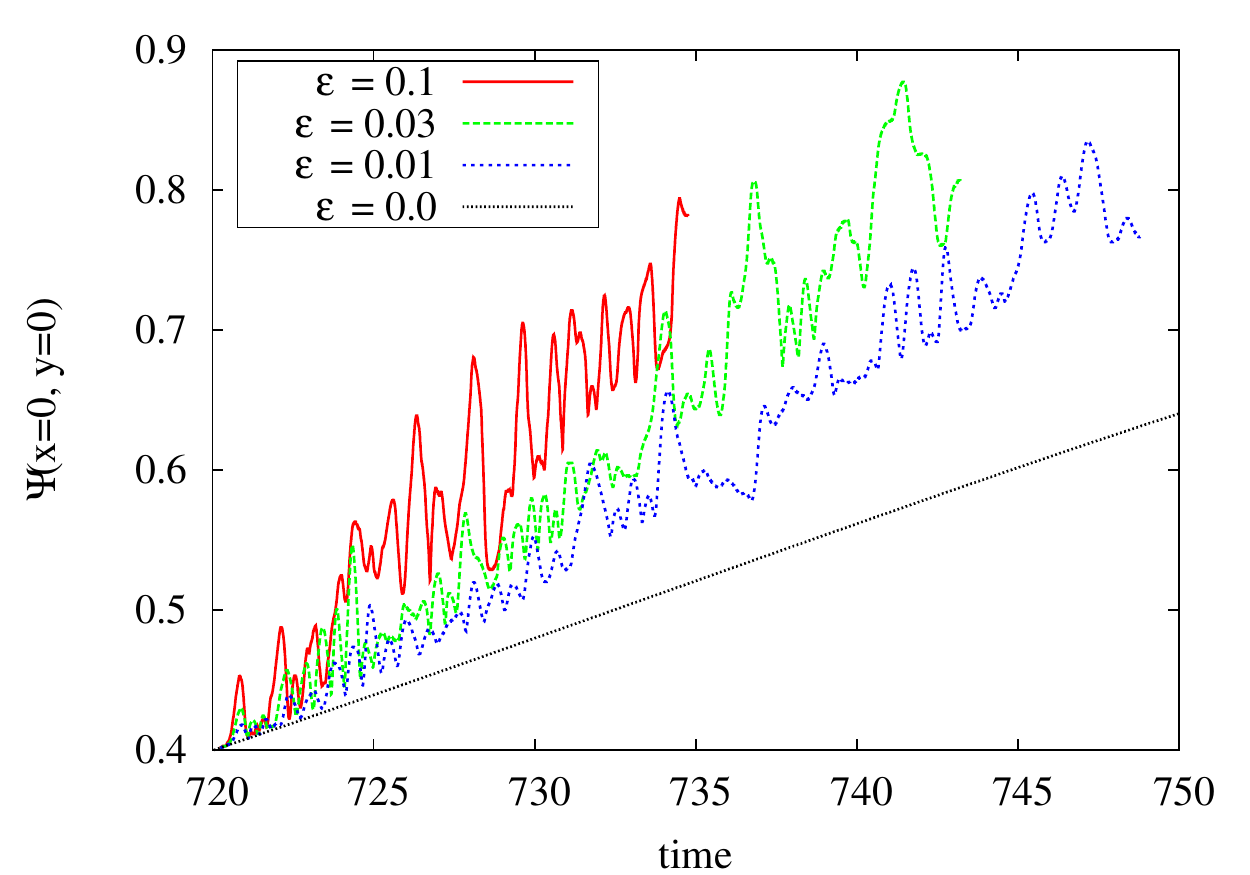}
		\caption{Selected time traces of the magnetic flux at the location of the laminar $X$-point ($x=0, y=0$), 
for $\eta=6.5\times10^{-5},k_f=10$ and different values of $\epsilon$.}
	\label{timetrace}
\end{figure}

A typical snapshot of these simulations is shown in \fig{global_snap}. 
The large-scale reconnecting configuration is manifest; the turbulent motions, which are also clearly visible, are not sufficiently strong to destroy it\footnote{There are regions with qualitatively different turbulence. This is because, whereas our turbulent forcing is statistically
homogeneous, the background magnetic and velocity fields associated
with the large-scale reconnecting configuration are not.}, so the reconnection problem is still well posed.
In the presence of turbulence, the current sheet is no longer straight as in the laminar case, but now wiggles about its original (i.e., laminar) location ($x=0$): see~\fig{contours}.
We find that secondary islands (plasmoids) are constantly present, being continuously generated and expelled from the sheet (whereas the laminar simulations at most display one plasmoid). The size of these plasmoids increases with $\epsilon$, as does the turbulent distortion of the sheet.\\ 
\indent We would like to measure the reconnection rate, $\Eeff$, in the presence of turbulence. In the laminar case, $\Eeff={\rm d} \psi_X/{\rm d} t$.
However, in turbulent simulations, the wandering of the current sheet makes the reconnection rate difficult to define. To quantify the effective reconnection rate, smoothed
over the rapid turbulent fluctuations, we employ the following procedure. Consider the representative timetraces of $\psi(x=0,y=0)$ of~\fig{timetrace}.
We see that they can be described by mostly upward fluctuations 
on top of a monotonically increasing baseline\footnote{Occasional downward fluctuations are due to plasmoids crossing the $(x=0,y=0)$ location.}.
We have checked that the minima of these curves correspond to maxima of the current. The reason is that, because of the turbulence, the actual very thin current sheet wanders around the original (laminar) $X$-point location --- see \fig{contours}. 
As it deviates from this location, the value of $\psi_X$ increases, since the laminar $\psi$ profile has a minimum at the $X$-point. When the current sheet crosses the location of the laminar $X$-point, the measured current is maximum, and $\psi$ is again minimum. 
This suggests that the slope of a linear fit to the local minima of these timetraces is an accurate diagnostic of the reconnection rate and that is what we use.
Error bars are estimated by reducing the interval of values of $\psi$ where the minima are selected by $0.05$ at both ends.
Although the turbulence is switched on at $\psi_X \approx 0.4$, we only apply this diagnostic for $\psi\gtrsim0.5$
so that the turbulence can reach a saturated stage before we assess its effect on the reconnection rate.
We emphasize that this is not the only possible diagnostic of the reconnection rate and indeed we have tried several different ones, with similar results.

\begin{figure}
	\centering
	\includegraphics[height=6cm]{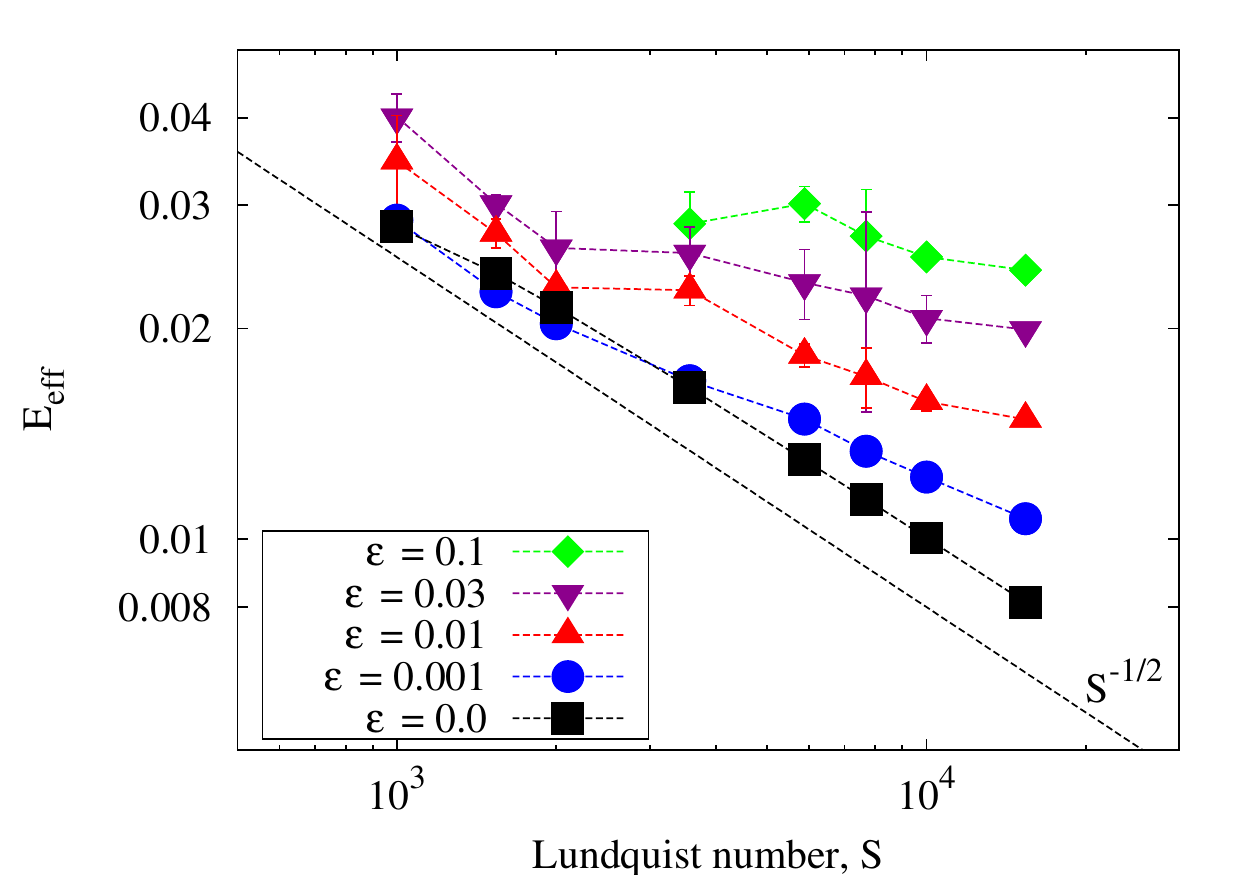}
	\caption{Plot of the effective reconnection rate (electric field) as a function of the Lundquist number $S=1/\eta$ for different values of $\epsilon$ and $k_f=10$. Dashed line shows $S^{-1/2}$ scaling.
Data for $\epsilon=0.1$ and $S\leq 3\times 10^3$ is not shown because the magnitude of the fluctuations and the relatively low values of $S$ prevent a statistically accurate computation of $\Eeff$.}
	\label{eta_scaling}
\end{figure}

The resulting dependence of the effective turbulent reconnection 
rate~$\Eeff$ on~$S=1/\eta$ is shown in~\fig{eta_scaling} for a range of values of the injected power~$\epsilon$.
For comparison, we also plot the reconnection rate obtained in the laminar runs, $E_{\rm SP}$, which yields very good agreement with the SP scaling of $S^{-1/2}$ (dashed line). 
%
%
The immediate observation, and the main result of this Letter, is that, 
\emph{at sufficiently large values of $S$,  all sequences with $\epsilon > 0.003$ exhibit a dependence on $S$ that is significantly shallower than the SP dependence of $S^{-1/2}$ and may even consistent with a $S$-independent value as $S\rightarrow\infty$}.
Also remarkable is that, common to all finite-$\epsilon$ sequences, there is a clear transition that does not seem to depend on $\epsilon$, always taking place around $S\approx 2\times10^{3}$.
Specifically, at values of $S$ below this, the reconnection rate in the turbulent runs is enhanced over the SP value, but only modestly, and still scales as $\sim S^{-1/2}$. Above this critical value turbulence has a much stronger effect on reconnection.
\begin{figure}
	\centering
		\includegraphics[height=6cm]{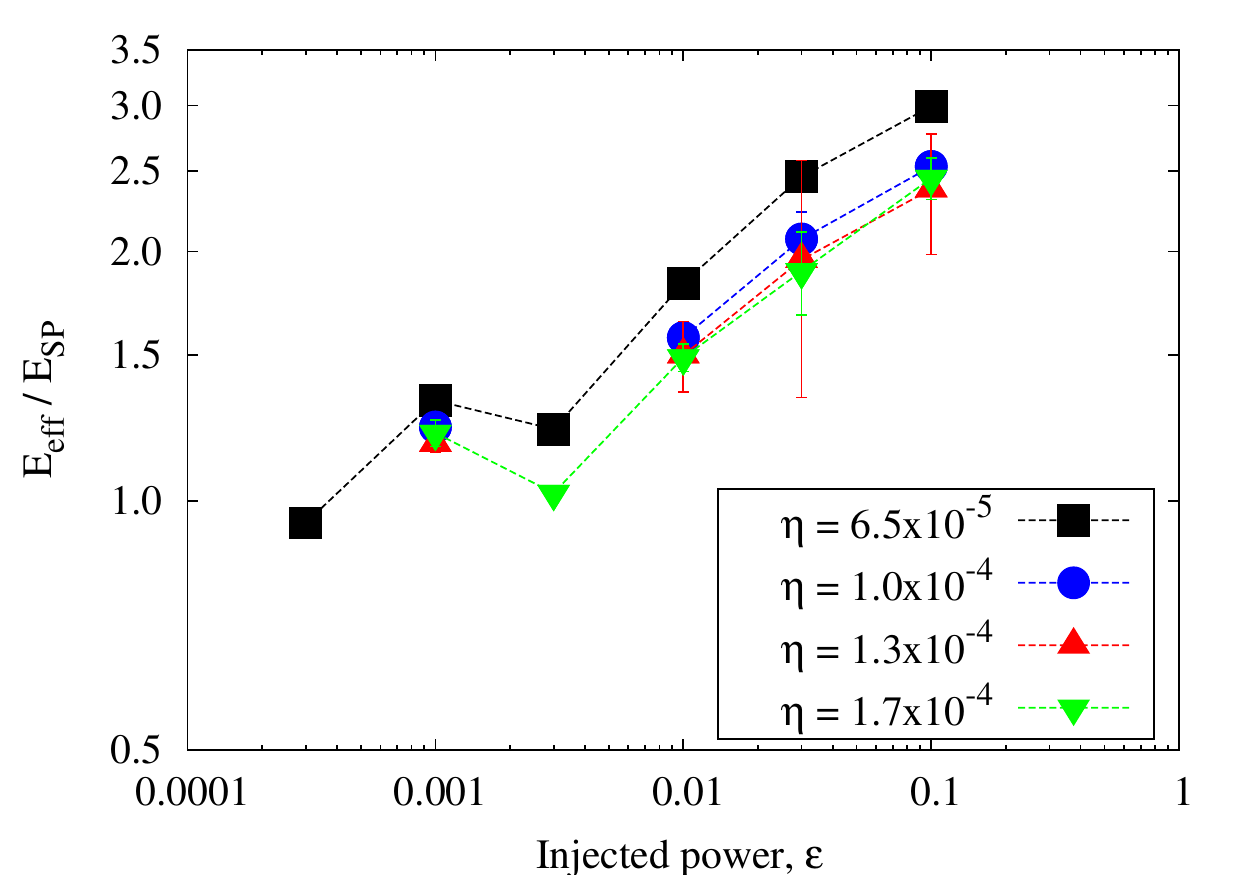}
	\caption{Plot of the reconnection rate enhancement factor as a function of the injected (turbulent) power, $\epsilon$, for the smallest values of $\eta$ and $k_f=10$.}
	\label{epsilon_scaling}
\end{figure}

Shown in~\fig{epsilon_scaling} is the dependence of the reconnection rate enhancement factor $\Eeff/E_{\rm SP}$ on the injected power $\epsilon$.
There is a distinct transition at $\epsilon\approx0.003$.
For smaller $\epsilon$, the dependence of the reconnection rate on $\epsilon$ is very weak, roughly consistent $\Eeff/E_{\rm {SP}} \sim 1$; for $\epsilon \gtrsim 0.003$, the enhancement factor increases with $\epsilon$ in a way which is consistent both with a power law $\Eeff/E_{\rm {SP}}\sim\epsilon^\alpha$, where $\alpha\approx 0.15-0.25$, and with a logarithmic dependence. Note that the power law exponent is significantly shallower than the LV prediction of $\epsilon^{1/2}$ for the 3D case.


\section{Discussion and Conclusion}
In summary, we have performed a detailed numerical study of the effect of background turbulence on 2D incompressible resistive magnetic reconnection. Our key finding is that, at sufficiently large values of the Lundquist number and moderate values of the injected (turbulent) power, the scaling of the reconnection rate with $S$ is \emph{much weaker} than the SP scaling of $S^{-1/2}$  and is, in fact, \emph{consistent with asymptotically fast reconnection} (independent of $S$; see~\fig{eta_scaling}). 
Other important conclusions are: (i) 
the transition from slow to fast reconnection happens at a fixed (i.e., independent of $\epsilon$) value of the Lundquist number, $S\approx2\times10^3$ and
(ii) the enhancement of the reconnection rate over its laminar (SP) value exhibits a clear threshold in the injected turbulent power, $\epsilon\approx0.003$, below which  turbulence has little influence on the reconnection rate and above which significant enhancement of the reconnection by turbulence starts --- see~\fig{epsilon_scaling}.
In our simulations, we find that the rms turbulent velocity $\urms$ (averaged over the simulation box), is closely described by $\urms \approx 2.44(\epsilon/k_f)^{1/3}$. Thus the
Alfv\'enic Mach number associated with the transition to fast reconnection is $M_A\equiv\urms/V_A\approx0.16$ 
(note that the dependence of our results on $k_f$ has not yet been tested; this can influence the numerical values of both the $S$ and the $\epsilon$ transitions).


Our results imply that a novel mechanism of turbulent enhancement of reconnection exists which is operative already in 2D. 
While our results support the basic claim by LV of turbulent enhancement of reconnection, 
we emphasize that, because they are 2D, they cannot be explained by \emph{any} present theory, including theirs. 
One speculative possibility is that the enhancement of the reconnection rate is related to the formation of multiple plasmoids~\citep{Loureiro_etal-2007, Loureiro_etal-2009, Samtaney_etal-2009}. This hypothesis is strengthened by the fact that the transition from slow to fast reconnection (see~\fig{eta_scaling}) is independent of $\epsilon$ and occurs at a value of $S$ consistent with the critical Lundquist number for plasmoid generation~\citep{Loureiro_etal-2005}. 
As $S$ increases, turbulence might facilitate frequent plasmoid formation by the local, transient, enhancement of the
tearing mode instability parameter $\Delta'$~\citep{FKR} (i.e, the local enhancement of the aspect ratio of the current sheet), but whether plasmoids can indeed provide the enhancement of the reconnection rate that we observe is not yet known.\\
\indent In this study, the effect of viscosity has not been addressed. 
Our modelling choice has been to keep the magnetic Prandtl 
number fixed and equal to $1$. 
This is not characteristic of most astrophysical systems, where either $Pm\gg1$ (e.g., warm interstellar and intracluster medium, as well as some accretion disks) or $Pm\ll1$ (e.g., stars, planets and liquid-metal laboratory dynamos)~\citep{Brandenburg-2005}. 
We have adopted $Pm=1$ to avoid additional complications and to present a clear demonstration of principle --- that turbulence \textit{can} accelerate reconnection in two dimensions.
Changing $Pm$ from $1$ will result in two effects: i) different viscous cutoff scale; however, 
since our results appear to depend only on the outer scale features of the turbulence (i.e. $\epsilon$, and possibly also $k_f$), we do not expect this effect to be significant; ii) different laminar solution, with a broader current sheet and slower outflows~\citep{Park_etal-1984}. This might affect the instability of the current sheet to plasmoids and could potentially impact on the reconnection rate in the presence of turbulence. Investigations of how Prandtl number (and forcing scale) might affect our results are left for future work.\\
\indent A final \textit{caveat} is that, in this work, we have neglected two-fluid effects, whose 
importance in reconnection is now widely recognized.
There are many astrophysical environments where these effects will doubtlessly
play an important role (e.g. solar flares); a future publication will assess their importance in turbulent reconnection.

\section*{Acknowledgments}
Simulations were done at UKAEA Culham, NCSA, Tigress (Princeton) and NERSC. We thank the CMPD for computational support. 
This work was partially funded 
by EPSRC
and by the European Commission under the contract of Association between 
EURATOM and UKAEA.
The views and opinions expressed herein do not necessarily 
reflect those of the European Commission.
D.A.U.\ was supported by NSF Grant\, PHY-0821899 (PFC: CMSO); he thanks
the Leverhulme Trust Network 
for Magnetized Plasma Turbulence for travel support. 
A.A.S.\ and T.A.Y.\ were supported by STFC.


\label{lastpage}
\end{document}